\documentclass[aps,prb,twocolumn,showpacs]{revtex4}

\usepackage{amsmath}
\usepackage{graphicx}

\newcommand{\be}{\begin{equation}}
\newcommand{\ee}{\end{equation}}
\newcommand{\mb}[1]{\mathbf{#1}}
\newcommand{\nn}{\nonumber}
\newcommand{\Journal}[4]{#1 \textbf{#2}, #3 (#4)}

\newcommand{\PRev}{Phys. Rev.}

\begin{document}

\title{Role of the Dzyaloshinskii-Moriya interaction in
multiferroic perovskites}

\author{I. A. Sergienko}
\affiliation{Condensed Matter Sciences Division, Oak Ridge
National Laboratory, Oak Ridge, TN 37831, USA}
\affiliation{Department of Physics, The University of Tennessee,
Knoxville, TN 37996, USA}

\author{E. Dagotto}
\affiliation{Condensed Matter Sciences Division, Oak Ridge
National Laboratory, Oak Ridge, TN 37831, USA}
\affiliation{Department of Physics, The University of Tennessee,
Knoxville, TN 37996, USA}

\begin{abstract}
With the perovskite multiferroic $R$MnO$_3$ ($R$ = Gd, Tb, Dy) as guidance,
we argue that the Dzyaloshinskii-Moriya interaction (DMI) provides the microscopic
mechanism for the coexistence and strong coupling between ferroelectricity and incommensurate magnetism.
We use Monte-Carlo simulations and zero temperature exact calculations
to study a model incorporating the double-exchange, superexchange, Jahn-Teller and DMI
terms. The phase diagram contains a multiferroic phase between A and E antiferromagnetic phases,
in excellent agreement with experiments.
\end{abstract}


\maketitle

\section{Introduction}

Several recent discoveries of unusually strong coupling between the ferroelectric (FE) and magnetic order
parameters have revived the interest in the magnetoelectric effect.\cite{Fiebig05}
Due to the possibility of easily controlling the electric properties using magnetic fields, a class of
compounds, in which the magnetic order is incommensurate (IC) with the lattice period, is
particularly interesting for future applications.\cite{Kimura03a, Hur04}
Surprisingly, this class of multiferroic materials includes compounds with very diverse crystallographic
structures: the perovskite $R$MnO$_3$ ($R$ = Gd, Tb, Dy),\cite{Kimura03a,Goto04,Noda05,Kimura05b,
Kenzel05,Arima05} orthorhombic $R$Mn$_2$O$_5$ ($R$ = Tb, Ho, Dy),\cite{Hur_more04,Chapon04,Higa04,
Kobayashi04,Blake05} hexagonal
Ba$_{0.5}$Sr$_{1.5}$Zn$_2$Fe$_{12}$O$_{22}$,\cite{Kimura05a} and
Kagom\'e-staircase Ni$_3$V$_2$O$_8$.\cite{Lawes05}
Unfortunately, the values of the electric polarization achieved so far are two orders of
magnitude smaller than those in traditional ferroelectrics. However, it is essential to theoretically
understand the new mechanism of magnetoelectric coupling.

Generally, certain types of magnetic order can lower the symmetry of the system to one of the polar groups,
which allows for ferroelectricity. This is a central argument of the existing phenomenological models.
\cite{Baryakhtar83,Lawes05,Kenzel05}
According to the recent experimental results in
Ba$_{0.5}$Sr$_{1.5}$Zn$_2$Fe$_{12}$O$_{22}$,\cite{Kimura05a} Ni$_3$V$_2$O$_8$,\cite{Lawes05}
and TbMnO$_3$,\cite{Kenzel05} helical magnetic structures are the most likely candidates to host
ferroelectricity. In addition, X-ray diffraction studies in a number of the above materials have revealed
that the modulated magnetic structure is accompanied by structural
modulation.\cite{Kimura03a,Higa04,Arima05}
It is, therefore, a natural assumption that lattice displacements actively participate in the formation
of the ferroelectric state as well, even though, the FE displacements have not been measured directly,
owing to their smallness ($\sim 10^{-3}$ \AA, as can be deduced from the value of the FE polarization
$\mb P$). This calls for theoretical microscopic models providing a mechanism by which the FE lattice
displacements are induced and coupled to the IC magnetic structure.

In this paper, we concentrate on the perovskite manganites since their experimental phase diagram has
been studied in much detail.\cite{Kimura05b,Kimura03b} We show that the Dzyaloshinksii-Moriya interaction
(DMI), linearly dependent on the displacements of the oxygen ions surrounding transition metal ions,
is an essential ingredient in the theory of the magnetoelectric effect in IC magnets.
In independent work,\cite{Katsura04} it is suggested that the DMI induces the polarization of the
electronic orbitals, without the involvement of the lattice degrees of freedom.
In our alternative scenario, the effect of the DMI is twofold: it induces the FE lattice displacements
and helps to stabilize helical magnetic structures at low temperature.

\begin{figure}
\vspace{0mm}
\includegraphics[width=65mm]{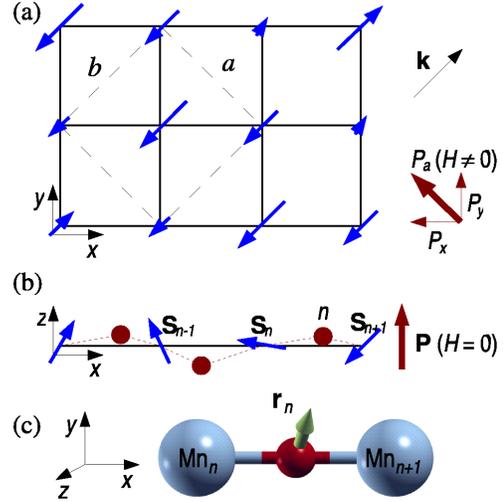}
\caption{\label{struct}%
(Color online)
(a, b) Sketch of the ground state structure of TbMnO$_3$. (a) Projection of the
Mn spins onto the $xy$-plane. Dashed lines are the boundaries of the unit cell in the orthorhombic
setting.
The diagram on the right illustrates the emergence of the in-plane component of $\mb P$
in applied magnetic field.
(b) The $zx$-projection of the spin structure and positions occupied by the O ions (filled circles).
(c) The Mn$_2$O ``molecule'' in the cubic perovskite structure.
The vector $\mb r_n$ denotes the displacement of the O ion.}
\end{figure}

\section{Semi-empirical discussion}

The cubic perovskite structure of $R$MnO$_3$ is orthorhombically distorted by the GdFeO$_3$-type
cooperative rotation of the MnO$_6$ octahedra. Here we use $x$, $y$, and $z$ to denote the directions in
the pseudo-cubic setting, and $a$, $b$, and $c$ are used for the directions in the orthorhombic
structure. The $c$-axis is parallel to $z$, and the relation between the in-plane axes is shown in
Fig.~\ref{struct}(a). In the following, we denote by ICM-FE the phase in which
the IC magnetic and FE order coexist.
In $R$MnO$_3$ ($R$ = Tb, Dy,\cite{Kimura03a,Goto04,Kimura05b} and, possibly,\cite{Noda05} Gd),
the ICM-FE phase is the ground state. The magnetic structure has the modulation vector
$\mb k$ along the $b$-axis within the $xy$-plane, and the planes are antiferromagnetically stacked
along the $z$-axis. Therefore, for a given Mn ion, the two nearest neighbors
in the positive directions of the $x$ and $y$ axes, have exactly the same spin, as shown
in Fig.~\ref{struct}(a). We use the index $n$ to enumerate the Mn ions along a chain in the
$x$-direction [Fig.~\ref{struct}(b)].
According to a neutron diffraction study of TbMnO$_3$,\cite{Kenzel05} the classical
low-temperature  spin structure of the Mn sublattice can be described as\cite{mnnote}
\be\label{spins}
S_n^i = S_0^i \cos(n\theta+\alpha_i),
\ee
where $i=x, y, z$, $S_0^x = S_0^y \approx S_0^z = 1.4$, and $\theta=0.28 \pi$.
The experiment was insensitive to the difference between $\alpha_x$ and $\alpha_z$,
but it was found that $\alpha_x = \alpha_y$. 
This structure is stable for temperatures $T<$~28~K. 
The lattice structure is also IC with the modulation vector $\mb k_l=2\, \mb k$.\cite{Kimura03a} 
The FE polarization $\mb P$ is oriented along the $z$-axis. 
For 28~K~$< T <$~41~K, another IC, collinear magnetic structure is found with no FE 
moment.\cite{Kimura03a,Kenzel05}

Considering the possible modifications of spin interactions by lattice displacements, we first examine the
isotropic superexchange. To address ferroelectricity, we assume that the positions of the Mn ions are
fixed, while the O ions can displace from their positions in the paraelectric phase.
We consider a chain of Mn ions in
the $x$-direction, and denote $\mb r_n = (x_n, y_n, z_n)$ the displacement of the O ion located
between the Mn spins $\mb S_n$ and $\mb S_{n+1}$ from its position in the idealized cubic perovskite
structure. In that structure, the Mn$_2$O ``molecule'' shown in Fig.~\ref{struct}(c) has inversion
symmetry. Since the symmetric superexchange ($\mb S_n\cdot \mb S_{n+1}$) is invariant under inversion
(\emph{i.~e.}~mere interchange of $\mb S_n$ and $\mb S_{n+1}$), this coupling can only
depend on even powers of $\mb r_n$. The axial symmetry of the Mn$_2$O ``molecule'' implies
\be
H_\text{ex} = - \sum_n \left[J_0 + \frac 1 2 J'_\parallel x_n^2 + \frac 1 2 J'_\perp (y_n^2+z_n^2)\right]
(\mb S_n \cdot\mb S_{n+1}),
\ee
where $J_0$, $J'_\parallel$, and $J'_\perp$ are constants, and the higher-order terms in $\mb r_n$ are neglected.

In the actual orthorhombically distorted structure, the oxygen ions are displaced, so that
\be
\mb r_n = (-1)^n \mb r_0 + \delta \mb r_n,
\ee
where $\mb r_0$ is constant ($r_0$ is a fraction of 1\AA),\cite{Blasco00} and $\delta \mb r_n$ is the additional
displacement, associated with the IC structure ($\delta r_n$ are of the order of $10^{-3}$ \AA, as mentioned
above). It can now be shown that the portion of the Hamiltonian depending on $\delta \mb r_n$ is
\begin{eqnarray}
\delta H_\text{ex} &=& \sum_n (-1)^{n+1} [J'_\parallel x_0 \delta x_n
+ J'_\perp (y_0 \delta y_n
\nn\\
&& +z_0 \delta z_n)](\mb S_n\cdot  \mb S_{n+1}) + H_\text{el},
\end{eqnarray}
where the elastic energy
\be
H_\text{el}=\frac \kappa 2 \sum_n (\delta x_n^2 + \delta y_n^2+ \delta z_n^2)
\ee
is assumed isotropic for clarity, the stiffness $\kappa>0$, and the second order terms in 
$\delta \mb r_n$ are neglected in the magnetoelastic term.
Minimizing $\delta H_\text{ex}$ with respect to the displacements and using~(\ref{spins}), we obtain
\be
\delta z_n =(-1)^n \frac {J'_\perp z_0}{2\kappa}
\sum_i S_0^{i2}\{\cos\theta+\cos[(2n+1)\theta+2\alpha_i]\},
\ee
and similar expressions for $\delta x_n$ and $\delta y_n$. Hence, this model reproduces the 
observed structural modulation with the wave vector $2\,\mb k$.\cite{Kimura03a} However, it is 
insufficient to explain the net FE polarization, since $\sum_n \mb r_n = 0$ exactly. 

On the contrary, the DMI, \emph{i.~e.} anisotropic exchange interaction\cite{Dzyalo58,Moriya60} 
$\mb S_n \times \mb S_{n+1}$ changes its sign under inversion. 
Using group theory, we find that the following 
expression is invariant under \emph{all} symmetry operations of the MnO$_2$ ``molecule'' 
in the perovskite structure:
$\mb D^{\mb a}(\mb r_n)\cdot[\mb S_n\times\mb S_{n+1}]$, where 
\be\label{Dvecs}
\mb D^{\mb x}(\mb r_n) = \gamma(0, -z_n, y_n), %\text{ and } 
\quad \mb D^{\mb y}(\mb r_n) = \gamma(z_n, 0, -x_n)
\ee
for the Mn-O-Mn bonds along the $x$ and $y$ axes, respectively.
The form of~(\ref{Dvecs}) can also be obtained\cite{Moskvin77} by perturbative calculations within the
Anderson-Moriya theory of superexchange.\cite{Moriya60} 
For the Mn chain in the $x$-direction, the portion of the Hamiltonian depending on $\delta \mb r_n$,
\be\label{hamDDMI}
\delta H_\text{DM} =\sum_n \mb D^{\mb x}(\delta\mb r_n)\cdot[\mb S_n\times\mb S_{n+1}]+ H_\text{el},
\ee
is minimized by
\be\label{displ}
\delta z_n=\frac \gamma \kappa S_0^x S_0^z \sin\theta\sin(\alpha_x-\alpha_z)
\ee
and $\delta x_n=\delta y_n=0$. The same result is obtained for the Mn-chain in the $y$-direction.
Hence, the displacements in this model do not depend on $n$, leading to 
a net FE polarization along the $z$ axis. % in zero magnetic field. 
Moreover, Eq.~(\ref{hamDDMI}) explains
why the IC phase with collinear spin structure is paraelectric. Clearly, the equilibrium value of
$\delta \mb r_n$ vanishes in this case. 

In applied magnetic fields, the magnetic structure changes through various phase 
transitions.\cite{Kimura05b}
Thus far, the magnetic structure was only determined experimentally for the magnetic field of 4 T 
oriented along 
the $a$ axis.\cite{Kenzel05} It is similar to the zero-field structure, and it does not induce the
in-plane component of $\mb P$.\cite{Kimura05b}
However, as can be seen from Eqs.~(\ref{Dvecs}) and~(\ref{hamDDMI}), the in-plane component of $\mb P$ 
can be induced, in general. It is plausible to assume that the Mn sites that have the same spin 
at zero field, also have the same spin in the applied \emph{uniform} fields [Fig.~\ref{struct}(a)]. 
Hence, the $x$-component of $\mb P$ is always equal in absolute value 
but opposite in sign to its $y$-component, and the in-plane polarization is always directed 
along the $a$ axis,\cite{polnote} in agreement with the experiments.\cite{Kimura03a,Kimura05b}

Equation~(\ref{displ}) allows us to estimate the value of $\gamma$. For typical phonon frequencies 
$\kappa \approx 1$eV/\AA$^2$, and we obtain $\gamma\approx 1$meV/\AA. Correspondingly,
the magnitude of $\mb D(\mb r_n)$ is of the order of $\gamma z_n\approx 0.1$ meV $\approx 1$ K,
which agrees well with the previous experimental estimations for perovskite 
manganites.\cite{Tovar99,Skumryev99,Deis02}

\section{Stabilization of the magnetic helix}

\begin{figure}
\includegraphics[clip]{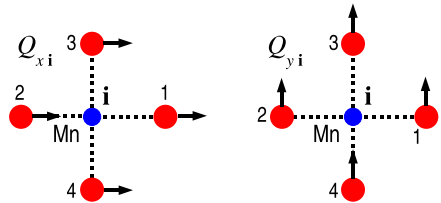}
\caption{\label{phonfig}
(Color online) Displacements of oxygen ions (large circles) corresponding to the
FE phonon modes in two dimensions.}
\end{figure}

To further understand the role of the DMI in the stabilization of the ICM-FE phase, we consider
the Hamiltonian
\begin{eqnarray}\label{ham}
H &=& -\sum_{\mb{i a} \alpha\beta\sigma} t^{\mb a}_{\alpha\beta} 
d^\dagger_{\mb i\alpha\sigma} d_{\mb i + \mb a \beta\sigma} - J_\text{H} \sum_{\mb i} 
\mb{s_i}\cdot \mb{S_i} \nn\\
&& + J_\text{AF} \sum_{\mb i \mb a} \mb{S_i}\cdot \mb{S}_{\mb i + \mb a}
+ \sum_{\mb i \mb a} \mb D^{\mb a}(\mb r)\cdot[\mb {S_i}\times \mb S_{\mb i + \mb a}]\nn\\
&& + \frac{\kappa_1}{2} \sum_{\mb i} (Q_{x\mb i}^2+Q_{y \mb i}^2) + 
H_\text{JT}+ \frac{\kappa_2}{2} \sum_{\mb i}\sum_{m} Q_{m \mb i}^2,\nn\\ 
\end{eqnarray}
based on the orbitally-degenerate double exchange model,\cite{Dagotto01} for a two-dimensional square 
lattice, representing a MnO$_2$ layer, with periodic boundary conditions and one $e_g$ electron per 
Mn$^{3+}$ ion. 
Here $d^\dagger_{\mb i\alpha\sigma}$ is the creation operator for the electron on site 
$\mb i$, orbital $\alpha = x^2-y^2 (\text{a}), 3z^2-r^2 (\text{b})$ and carrying spin 
$\sigma$. The hopping integrals are given by $t^{\mb x}_\text{aa}=
-\sqrt{3}t^{\mb x}_\text{ab}=-\sqrt{3}t^{\mb x}_\text{ba}=3 t^{\mb x}_\text{bb}\equiv t$ and
$t^{\mb y}_\text{aa}=\sqrt{3}t^{\mb y}_\text{ab}=\sqrt{3}t^{\mb y}_\text{ba} = 
3t^{\mb y}_\text{bb} = t$. Hereafter, $t$ is taken as the energy unit.
$J_\text{H}$ is the Hund's coupling constant between the $e_g$ electrons with spin 
$\mb s_{\mb i} = \sum_{\alpha\sigma\sigma'}d^\dagger_{\mb i\alpha\sigma} 
\boldsymbol{\sigma}_{\sigma\sigma'} d_{\mb i \alpha\sigma'}$ ($\boldsymbol{\sigma} = $ Pauli
matrices) and three $t_{2g}$ electrons treated as a classical three-dimensional spin $\mb{S_i}$.
In the following, $J_\text{H}$ is assumed to be infinite. $J_\text{AF}>0$ is the isotropic 
superexchange constant between the $t_{2g}$ spins. 
For simplicity, we consider DMI only between the $t_{2g}$ spins and assume that O ions can 
only move within the plane. $Q_{m \mb i}$ denote the classical phonon coordinates associated with 
the displacements of the four O atoms surrounding the Mn site $\mb i$. The doubly degenerate FE mode 
$(Q_{x\mb i}, Q_{y \mb i})$ is shown in Fig.~\ref{phonfig}. 
%given by $Q_x=(x_1+x_2+x_3+x_4)/2$ and $Q_y=(y_1+y_2+y_3+y_4)/2$, where the site index is omitted for 
%clarity. 
The Jahn-Teller (JT) interaction term is defined as usual,\cite{Dagotto01,Hotta03}
\be\label{hamJT}
H_\text{JT}=\lambda(q_{1\mb i}\rho_{\mb i} + q_{2\mb i} \tau_{x\mb i} + q_{3\mb i} \tau_{z\mb i})
+\frac 1 2 \sum_{\mb i} (2 q_{1\mb i}^2 + q_{2\mb i}^2  + q_{3\mb i}^2),
\ee
where $\rho_{\mb i}=\sum_{\alpha\sigma} d^\dagger_{\mb i\alpha\sigma}d_{\mb i\alpha\sigma}$,
$\boldsymbol{\tau}_{\mb i}=\sum_{\alpha\beta\sigma}d^\dagger_{\mb i\alpha\sigma} 
\boldsymbol{\sigma}_{\alpha\beta} d_{\mb i \beta\sigma}$ is the orbital pseudospin. The three-dimensional
JT phonon modes $q_{1,2,3 \mb i}$ have been appropriately redefined for our model in terms of the two dimensional
phonons $Q_{m \mb i}$.

The remnant phonon modes (not FE or JT) are assumed to have the same spring constant
$\kappa_2$, an assumption that can be easily removed if necessary. 
If $\kappa_1 =\kappa_2$, the ICM-FE phase is degenerate with other spin-canted structures. 
We assume $\kappa_1 <\kappa_2$ in order to lift the degeneracy in favor of the ICM-FE phase.

\begin{figure}
\begin{minipage}[t]{1.0\columnwidth}
\vspace{0pt}
\includegraphics[clip,width=86mm]{ZeroT1.eps}
\end{minipage}\\[10pt]
\begin{minipage}[t]{1.0\columnwidth}
\begin{center}
\includegraphics[clip,width=70mm]{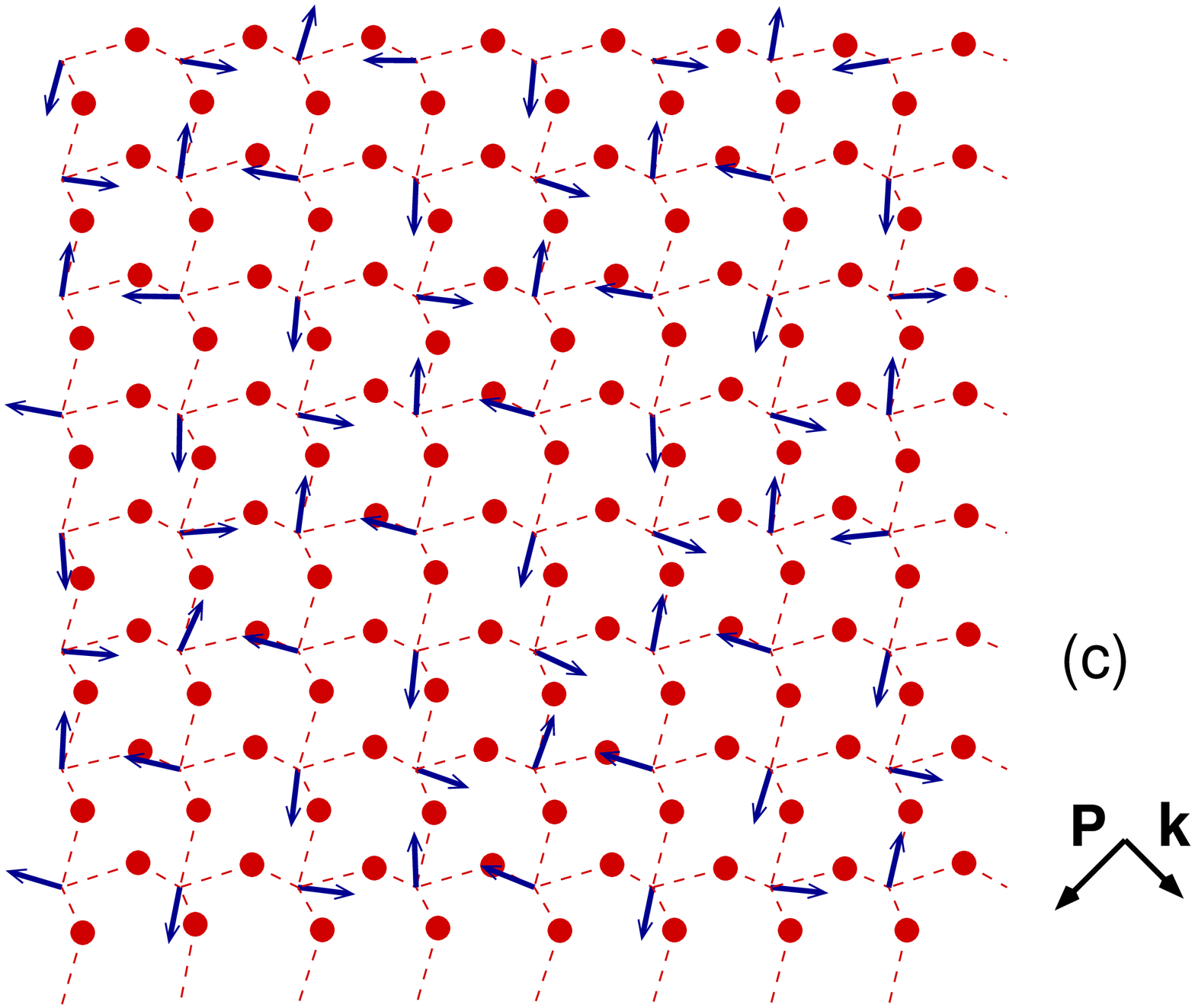}
\end{center}
\end{minipage}
\caption{\label{DEfig} 
(Color online)
Low temperature results for model~(\ref{ham}) with $\kappa_1=1$, $\kappa_2=10$.
(a) Ground state phase diagram. Broken lines are the MC results for the $8\times 8$ cluster.
Solid lines are the results of the calculations in the thermodynamic limit.
(b) Typical MC electronic density of states (DOS) as a function of the electronic energy $\omega$
in the ICM-FE phase.
(c) Typical MC snapshot in the ICM-FE phase. 
Shown are the ICM modulation vector $\mb k$ and polarization $\mb P$.}
\end{figure}

We have performed extensive Monte-Carlo (MC) simulations of $H$ with $\lambda = 0$ on an $8\times 8$ 
cluster, which is sufficiently large for our purposes. 
The low-temperature ($T = 0.01$) results are presented in Fig.~\ref{DEfig}. For small $\gamma$, the 
$J_\text{AF}$--$\gamma$ phase diagram incorporates the FM phase (which is the two-dimensional
precursor of the three-dimensional A-type antiferromagnetic phase) and the E-type phase,
in agreement with previous results for smaller clusters.\cite{Hotta03}
We find that the FM phase is represented by two states, separated by the dotted line in  
Fig.~\ref{DEfig}(a). The regular FM phase with all spins aligned in the same direction is stable on the 
left hand side of the dotted line, while the right hand side is the domain of stability of the so-called 
``twisted'' phase with the period equal to the cluster length. 
This is a well-known effect of the finite cluster size and 
periodic boundary conditions.\cite{Kubo82,Zang97,Dagotto98} 

Our most important numerical result is the stabilization of a new modulated (ICM-FE) phase at 
finite $\gamma$. The presence of the three phases in the ground state phase diagram 
is in excellent agreement with the experiments.\cite{Kimura03a,Kimura03b}  The period of the ICM-FE phase
is fixed in the MC simulations due to the small cluster, 
but it is expected to vary smoothly with the parameters of $H$ when the lattice size is increased,
as discussed below. 
A typical snapshot showing the spin structure and O displacements 
in this phase is presented in Fig.~\ref{DEfig}(c)
showing the spin modulation with $\mb k =(\pi/2,\pi/2)$. In the two-dimensional model studied here, 
$\mb P$ as well as the spins lie within the plane. 
This is a result of confining the O ions to move within the 
plane. The ground state with $\mb P$ along the $z$-direction can be obtained if model~(\ref{ham}) is
considered in three dimensions and the parameters of the model are changed in accordance with the
orthorhombic distortions of the cubic lattice to remove the equivalence of the crystallographic directions.

Figure~\ref{DEfig}(b) shows the calculated electronic density of states in the ICM-FE phase, with a gap
at the chemical potential $\mu$. This demonstrates the insulating character of the new
phase, which insures that the polarization is not screened out by the free charge 
carriers. 

To elucidate the influence of the finite size effects on the ground state phase diagram, 
we calculated the energies for the idealized spin and phonon configurations, for which 
the $e_g$ part of $H$ can be diagonalized exactly in the thermodynamic limit. 
This calculation becomes exact at zero temperature for the model considered here, with
adiabatic phonons and classical spins, under the reasonable assumption that no other
phases appear as ground states. In particular,  we obtain for the ICM-FE phase,
\be\label{EICM}
E_\text{ICM-FE} = N(-1.60 \cos \frac\theta 2 + 2 J_\text{AF}\cos \theta -\frac{\gamma^2}{4\kappa_1}
\sin^2\theta),
\ee
where $N$ is the number of Mn sites in the lattice and $\theta$ is the angle between two neighboring
spins. As can be seen from Fig.~\ref{DEfig}(a), while the boundary between the FM and E phases remains 
practically unchanged, the increase of the lattice size strongly affects the ICM-FE phase. This is 
expected, since $\theta$ can now assume the continuous values found by the minimization of~(\ref{EICM}),
instead of being restricted by the boundary conditions. 
Generally, the absolute value of $\mb k$ changes from 
zero at the FM-ICM phase transition to some finite value at the discontinuous ICM-E phase transition.
Also, in the phase diagram obtained by this procedure, only the regular FM paraelectric phase is found 
for small $\gamma$ and $J_\text{AF}<0.18$.

\begin{figure}
\includegraphics[clip,width=80mm]{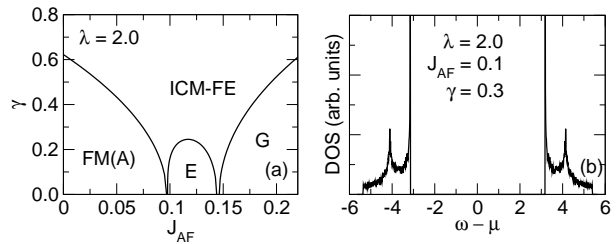}
\caption{\label{lam_fig} 
Results of exact diagnolization of Hamiltonian~(\ref{ham}) at T=0 in the thermodynamic limit for $\lambda=2.0$,
$\kappa_1=1$, and $\kappa_2=10$.
(a) Ground state phase diagram, (b) Electronic density of states in the ICM-FE phase.}
\end{figure}

The minimal value of $\gamma$ needed to stabilize the ICM-FE phase between A and E phases is calculated
in the thermodynamic limit to be about 0.3 [see Fig.~\ref{DEfig}(a)], which corresponds to approximately 
200 meV/\AA\ in the physical units. 
This is two orders of magnitude larger than the expected empirical value obtained in Sec. II. 
However, the numbers can change as the model is improved. In particular, our 
zero temperature results demonstrate that the minimal $\gamma$ needed can be made 
infinitesimal if the JT interaction is taken into account. In Fig.~\ref{lam_fig} we show the results for 
a realistic value of $\lambda = 2.0$. The JT interaction also significantly improves insulating properties of
the model. The details of this study for finite $\lambda$, 
including the results on orbital order and MC simulations, will be reported elsewhere.

\begin{figure}
\includegraphics[clip,width=80mm]{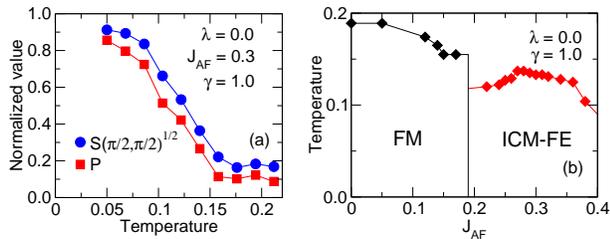}
\caption{\label{Tfig} (Color online)
Finite temperature MC results for Hamiltonian $H$; 
$\kappa_1=1$, $\kappa_2=10$.
(a) Typical temperature dependence of the ICM and FE order parameters. The order 
parameters are normalized to their maximum value at $T=0$.
(b) Temperature phase diagram. }
\end{figure}

The results of the finite temperature MC study are reported in Fig.~\ref{Tfig}.
Figure~\ref{Tfig}(a) depicts a typical temperature dependence of the ICM structure factor and  
absolute value of $\mb P$. Within the simulation errors, the two order parameters have the same
ordering temperature, demonstrating the strong mutual influence of ferroelectricity and 
magnetic order. The corresponding temperature phase diagram is presented in Fig.~\ref{Tfig}(b). 
Here, we find only one ICM phase with a direct phase transition from the disordered state. In order to
account for the collinear ICM phase at the intermediate temperatures, one has to include in the model
the anisotropy associated with the orthorhombic distortions. Then, different crystallographic axes become 
inequivalent
and the correlations of the corresponding spin components become finite at different 
temperatures, due to magnetic anisotropy.\cite{Nagamiya67}
 
\section{Conclusions}

We have demonstrated that the DMI 
provides a natural explanation for
the coexistence and strong coupling between ferroelectricity and IC magnetism in the perovskite manganites.
The empirically estimated value of $\gamma$ is in agreement with previous experiments.
The ICM-FE phase appears in the nearest-neighbor interaction model~(\ref{ham}) as a result of the 
competition between 
the double exchange, superexchange and DMI. 
Therefore, our model favors a mechanism of 
stabilization of helical magnetic order which is qualitatively different from the theories based on spin 
frustration.\cite{Blake05, Kimura03b,Nagamiya67} 
Unlike in the model suggested in Ref.~\onlinecite{Gennes60}
the noncollinear spin state is realized in the \emph{undoped} regime, and thus it is stable with respect
to phase separation.\cite{Dagotto01}

We believe that, due to its universality, the DMI is also relevant in other IC multiferroics.
Indeed, the DMI is independent of the orbital structure, it only involves the interaction between
spins induced by the symmetry-breaking (in particular FE) ionic displacements. 
In support of this assertion, we note that, despite the diversity of crystal structures, 
$\mb P \perp \mb k$ \emph{always} for all the IC multiferroics discovered so far, with or without applied 
magnetic field,\cite{Kimura03a,Noda05,Kimura05b,Kenzel05,Hur04,Blake05,Kimura05a,Lawes05} which is 
naturally explained by the DMI.

\begin{acknowledgments}
We thank S. H. Curnoe, T. Egami, T. Kimura, D. Mandrus, A. Moreo, D. J. Singh, and A. I. Zheludev 
for useful discussions. Part of the code for the MC simulations was adopted from SPF program developed 
by G. Alvarez (http://mri-fre.ornl.gov/spf). Figure~\ref{struct}(c) was generated using XCrySDen program 
(http://www.xcrysden.org).
I.S. was supported by NSF DMR-0072998. E.D. was supported by NSF DMR-0443144. Research at 
Oak Ridge National Laboratory is sponsored by the Division of Materials Sciences and Engineering, 
U.S. Department of Energy, under contract DE-AC05-00OR22725 with UT-Battelle, LLC.
\end{acknowledgments}


\begin{thebibliography}{00}
\bibitem{mnnote} Here, we do not consider the magnetic moments of the rare-earth ions explicitly,
since they are not directly related to the FE order.
\end{thebibliography}
\end{document}